\documentclass[preprint,showpacs,preprintnumbers,amsmath,amssymb]{revtex4-1}
\usepackage{graphicx}
\usepackage{rotating}
\begin{document}

\renewcommand{\theequation}{\thesection.\arabic{equation}}

\newcommand{\re}{\mathop{\mathrm{Re}}}

\newcommand{\be}{\begin{equation}}
\newcommand{\ee}{\end{equation}}
\newcommand{\bea}{\begin{eqnarray}}
\newcommand{\eea}{\end{eqnarray}}

\title{Are singularities the limits of cosmology?}

\author{Mariusz P. D\c{a}browski}
\email{mpdabfz@wmf.univ.szczecin.pl}
\affiliation{\it Institute of Physics, University of Szczecin, Wielkopolska 15, 70-451 Szczecin, Poland}
\affiliation{\it Copernicus Center for Interdisciplinary Studies,
S{\l }awkowska 17, 31-016 Krak\'ow, Poland}

%\date{\today}

\input epsf

\begin{abstract}
We refer to the classic definition of a singularity in Einstein's general relativity (based on geodesic incompletness) as well as to some other criteria to evaluate the nature of singularities in cosmology. We review what different (non-Big-Bang) types of singularities are possible even in the simplest cosmological framework of Friedmann cosmology. We also show that various cosmological singularities may be removed or changed due to the variability of physical constants.

\end{abstract}

%\pacs{98.80.-k; 98.80.Jk; 04.20.Dw; 04.20.Jb}

\maketitle

\section{What are singularities?}
\label{intro}
\setcounter{equation}{0}

Asking about the limits in cosmology is almost the same as asking about singularities. They are one of the most intriguing objects since they open the way to the new physics -- the physics which cannot be described by actual theories of the universe. The singularities are just some infinities of the physical and mathematical quantities which lead to experimental or observational problem since one cannot practically measure any infinite quantity with any type of a realistic device. This is why we say that our (whatever) the theory, once it possesses a singularity. However, the singularities appear in physics and are formally described by mathematics so that we have to somehow deal with them. In fact, they appear in all physical theories. For example, in Newton's theory of gravity there is a singularity when a spherical shell collapses to a point at its center. It is obvious for cosmologists that we also experience singularities in Einstein's gravity and best-known example is a Big-Bang singularity (corresponding to the beginning of the evolution of the universe) and a black hole singularity (corresponding to the collapsed matter as a result of gravitational attraction).

In some intuitive approach we can talk about singularity as a {\it ``place''} in which some kind of a {\it ``pathology''} is observed. We know that physical fields, such as  for example the electric field, can be singular at the place where a point charge is situated. This is a physical field singularity which resides in space. However, in Einstein's gravity we deal with space itself and a singularity means that we cannot even talk about the ``background'' space as being something measurable. Then, there is really a big  difference between physical field singularities and spacetime singularities and they are quite independent (a well-known example is from string theory in which both gravitational and strong coupling of the field singularities are present).

Various definitions of singularities have been proposed in the literature. Some of them are based on the blow-up (infinity) of the curvature tensor and curvature invariants, differentiating them as ``boundaries'' of spacetime, but there are some ``pathological'' examples such as gravitational waves, conical singularities which do not fall into the scheme. In general, the matter is very subtle and so far the best definition is considered to be the definition based on the condition of {\it geodesic incompletness} (e.g. \cite{HE,wald}) which allows to practically detect them without ``adopting'' them into the theory. According to this definition {\it spacetime is singular}, if there exists {\it at least one geodesic which is incomplete} i.e. which cannot be extended in at least one direction and has only a finite range of an affine parameter (which is a proper time of an observer or length of a curve for non-null geodesics).

This definition seems to give {\it the limits} of gravity and especially of cosmology. ``Points'', ``regions'', ``holes'' which are not reachable by our physical theory are suspected to be singularities of spacetime. The requirement of geodesic incompleteness seem to be a kind of ``minimalistic'' approach to the problem which in fact {\it does not tell us anything about the nature of the singularities} (e.g. how they impact on physical and geometrical quantities).

In this article we will try to get into some more subtlety of the matter applying other definitions in order to differentiate various singularities. One of the important points is to answer the question, whether there exist theories which are singularity-free, and whether there is only one type of singularities or there are more of them, and since this is really the case, then to investigate which of them are the most dangerous for the matter/an observer to approach.

A lot of generalized theories of gravity have been proposed which tried to avoid a Big-Bang singularity. Among them are the superstring and the brane theories \cite{string,brane}, loop quantum gravity \cite{LQC}, higher-order gravity \cite{f(R)}, and many others. One of the achievements of such theories was the extension of the evolution of the universe through a big-bang singularity like in the pre-big-bang \cite{PBB} and the cyclic \cite{cyclic} scenarios.

The motivation for our discussion of singularities as limits of cosmology comes mainly from the discovery of the accelerated expansion of the universe  \cite{supernovaeold,supernovaesupernew}. After this discovery some deeper studies of the phenomenon of dark energy which drives accelerated expansion showed the plethora of new singularities, now known as ``exotic'' singularities, which are quite different from the standard Big-Bang \cite{nojiri}.

\section{Big-Bang and non-Big-Bang singularities in cosmology.}
\setcounter{equation}{0}
\label{models}

First example of a non-Big-Bang type (from now on type 0) of a singularity in cosmology which is compatible with observations is a Big-Rip (BR or type I) associated with phantom dark energy \cite{phantom}. Another examples are: a sudden future singularity (SFS or type II) \cite{SFS}, generalized sudden future singularities (GSFS), finite scale factor singularities (FSF or type III),  big-separation singularities (BS or type IV) \cite{nojiri}, and $w$-singularities (type V) \cite{wsing}. There are also some versions of a Big-Rip known as: a Little-Rip \cite{LRip} and a Pseudo-Rip \cite{PRip}. It seems fascinating that some of these singularities are ``weaker'' and some are  ``stronger'' than a Big-Bang in the sense of leading to infinities of some specific type and not to the other type.

In order to understand what we mean by ``weaker'' and ``stronger'' we have to refer to some mathematical tools to investigate the problem.

\subsection{The strength of singularities}

One of the methods to study the nature of non-Big-Bang singularities is to apply two definitions which measure their ``strength''. According to the definition of Tipler \cite{tipler} a singularity is strong if an extended object represented by three linearly independent, vorticity-free geodesic deviation vectors at $p$ parallely transported along causal geodesic $l$ is {\it crushed to zero volume} at the singularity by infinite tidal forces. In mathematical terms it means that at least one component of the tensor $I_j^i (\tau) = \int_0^{\tau} d\tau' \int_0^{\tau'} d\tau'' |R^i_{~ajb}u^a u^b|$ ($R^i_{~ajb}$ - the Riemann tensor, $u^a$ - four-velocity vector, $a,b,i,j = 0,1,2,3$, $\tau$ - proper time) diverges on the approach to a singularity at $\tau = \tau_s$. On the other hand, according to the definition of Kr\'olak \cite{krolak}, a singularity is strong if the {\it expansion} of every future-directed congruence of null (timelike) geodesics emanating from the point $p$ and containing $l$ {\it becomes negative} somewhere on $l$ or, in mathematical terms, if at least one component of the tensor $I_j^i(\tau) = \int_0^{\tau} d\tau' |R^i_{~ajb}u^a u^b|$ diverges on the approach to a singularity at $\tau = \tau_s$. For the null geodesics one replaces Riemann tensor by the Ricci tensor components \cite{HE}.

\subsection{Geodesics and geodesic deviation}

The non-Big-Bang singularities can also be studied in the context of the particles worldlines and a change of the distance between these wordlines. In order to study geodesic incompleteness one should use geodesic equation \cite{HE}
\be
\label{geodesic}
\frac{d^2x^{a}}{d\lambda^2} + \Gamma_{bc}^{a} \frac{dx^{b}}{d\lambda} \frac{dx^{c}}{d\lambda} = 0~,
\ee
where $\Gamma_{bc}^{a}$ are the Christoffel connection coefficients, $x^a$ - the coordinates, and $\lambda$ is an affine parameter, and the geodesic deviation equation \cite{HE}
\bea
\label{deviation}
\frac{D^2n^{a}}{d\lambda^2} + R^{a}_{~bcd} u^{b} n^{c} u^{d} = 0~,
\eea
with $n^{a}$ being the deviation vector separating neighboring geodesics (particles worldlines) which describes the propagation of the distance between geodesics.

\subsection{Spacetime averaging}

It is also useful to use the notion of spacetime averaging of singularities due to Linde and Raychaudhuri \cite{averaging}. The idea is that one may {\it average physical and kinematical scalars $\chi$} over the whole open spacetime (provided the scalars vanish rapidly at spatial and temporal infinity) as follows
\be
\label{aver}
<\chi> = \lim_{x^a \to \infty} \frac{\int\int\int\int_{-x^a}^{x^a} \chi \sqrt{-g} d^4 x}{\int\int\int\int_{-x^a}^{x^a} \sqrt{-g} d^4 x}~~,
\ee
where $g$ is the determinant of the spacetime metric tensor. By an open model it is meant that the ratio of the 3-volume hypersurfaces to a 4-volume of spacetime vanishes, i.e.,
\be
\frac{\int\int\int \sqrt{\mid ^3g \mid} d^3x}{\int\int\int\int \sqrt{-g}d^4x} = 0~~,
\ee
where $^3g$ is the determinant of the spatial part of the metric tensor. The vanishing of the average $<\chi>$ was supposed to be related to singularity avoidance in cosmology \cite{averaging}.

\subsection{Energy conditions}

One way to characterize the nature of spacetime singularities is to check if they obey the energy conditions \cite{HE}. The S(trong) E(nergy) C(ondition) tells us that  gravity is attractive and for the Friedmann isotropic cosmology reads as
\bea
\label{strong0}
\varrho c^2 + 3p \geq 0, \hspace{0.5cm} \varrho c^2 + p \geq 0~,
\eea
where $\rho$ is the mass density, $p$ is the pressure, and $c$ is the speed of light. The N(ull) E(nergy) C(ondition) is obeyed, if the flux of matter is non-spacelike and reads as
\be
\label{null}
\varrho c^2 + p \geq 0~.
\ee
The W(eak) E(nergy) C(ondition) guarantees positivity of energy:
\be
\label{weak}
\varrho c^2 + p \geq 0, \hspace{0.5cm} \rho c^2 \geq 0~,
\ee
and the D(ominant) E(nergy) C(ondition) requires that the pressure $p$ is smaller than the mass density $\rho$
\be
\label{dom}
\mid p \mid \leq \varrho c^2, \hspace{0.5cm} \varrho c^2 \geq 0~.
\ee

All the non-Big-Bang singularities are characterized by violation of all, some, or none of the energy conditions which is related to a blow-up of all, some, or none of the appropriate physical quantities such as: the scale factor, the mass density, the pressure, the barotropic index etc.

\section{Properties and classification of singularities.}
\setcounter{equation}{0}
\label{class}

%\subsection{No geodesic incompletness - no limits of cosmology.}

The explicit application of geodesic equation (\ref{geodesic}) to Friedmann cosmology
\bea
\left(\frac{dt}{d\tau}\right)^2 &=& A + \frac{P^2 +
kL^2}{a^2(t)}~,\\
\frac{dr}{d\tau} &=& \frac{P_1 cos{\phi} + P_2
\sin{\phi}}{a^2(t)} \sqrt{1-kr^2}~,\\
\frac{d\phi}{d\tau} &=& \frac{L}{a^2(t)r^2}~.
\eea
($A,P,L,P_1,P_2=$ const., $t,r,\phi$ - metric coordinates, $a(t)$ - the scale factor, $k = 0, \pm 1$ - curvature index) shows that some non-BB singularities can be continued through since geodesics {\it do not feel} them at all -- they are not singular there since for example $a_s = a(t_s)=$ const. at $t_s$ being the time of singularity, and there is {\it no geodesic incompletness} \cite{lazkoz}. However, the geodesic deviation equation (\ref{deviation}) (which measures the behavior of a bunch of geodesics) does feel singularities since at $t=t_s$ the Riemann tensor $R^{a}_{~bcd} \to \infty$.

As an example we see that with an SFS it is possible to ``go through'' singularity since we have
\be
R^{a}_{~0b0} = - \frac{\ddot{a}}{a} \delta_{b}^{a}, \hspace{0.5cm} (...)^{\cdot} = \frac{\partial}{\partial t},
\ee
($\delta_{b}^{a}$ is the Kronecker delta) which for integral curves of $u = \partial/\partial t$ (geodesics with an affine parameter t) gives
\be
\dot{u}^{a} = - R^{a}_{~0b0} n^{b} \propto \ddot{a}
\ee
which diverges to minus infinity at $t=t_s$.

Physically, it means that the tidal forces which manifest here as the (infinite) impulse which reverses (or stops) the increase of separation of geodesics and the geodesics themselves can evolve further -- the universe can continue its evolution through a singularity. This behavior is like a turning point of a harmonic oscillator.

A bigger surprise follows if one considers an extended object such as a classical string (here of a Polyakov type \cite{string})
\be
S = - \frac{T}{2} \int d\tau d\sigma \eta^{\mu\nu}g_{ab} \partial_{\mu}
X^{a} \partial_{\nu} X^{b}~~,
\ee
($T$ - string tension, $\tau, \sigma$ - worldsheet of the string coordinates, $\eta^{\mu\nu}$ - worldsheet metric, $\mu,\nu = 0,1$, $g_{ab}$ - spacetime metric) falling onto a non-BB singularity \cite{adam}. The point is that an invariant string size $S(\tau) = 2 \pi a(\eta(\tau)) R(\tau)$ (assuming circular ansatz with radius $R$)
shows that they are: infinitely stretched $S \to \infty$ at a Big-Rip (a string is destroyed) while at SFS the scale factor is finite at $\eta$-time so that
the invariant string size is also finite. The same is true for FSF, BS, and GSFS. This means that strings are not destroyed at such singularities.

The ``weakness'' of these singularities means that they do not exhibit the geodesic incompleteness and so the particles \cite{lazkoz} and even extended objects \cite{adam} may pass through them. Then, they are not ``dangerous'' and this is why they may appear in the very near future (e.g. about 10 mln years for SFS) \cite{SFSobserv,FSFobserv}.

Now, one can use the Puiseux series expansion for the scale factor \cite{LFJ2007}
\be
a(t) = c_0 + (t_s - t)^{\eta_0} + c_1 (t_s - t)^{\eta_1} + c_2 (t_s - t)^{\eta_2} + \ldots \hspace{1.cm} \eta_0 < \eta_1 < \ldots \hspace{0.3cm} c_0 > 0
\ee
in order to check the geodesic incompletness and the strength of non-BB singularities using geodesic equations and the definitions of Tipler (T) and Kr\'olak (K).
The summary of such investigations is given in table \ref{classif}. Here we can make the claim that if there is no geodesic incompleteness, then there are no limits to cosmology.

\begin{table*}
\caption{Classification of singularities in Friedmann cosmology. Here $t_s$ is the time when a singularity appears, $w=p/\rho$ is the barotropic index, $T$ - Tipler's definition, $K$ - Kr\'olak definition.}
\label{classif}
\begin{center}
\begin{tabular}{lcccccccccc}
\hline
\\
Type & Name & $t_s$ & a(t$_s$) & $\varrho(t_s)$ & p(t$_s$) & $\dot{p}(t_s)$ etc.  & w(t$_s$) & T & K\\
\hline
\\
0  & Big-Bang (BB) & $ 0 $ & $ 0$ & $\infty$ & $\infty$ &$\infty$& finite & strong & strong\\
I  & Big-Rip (BR) & $t_s $ &$\infty$ & $\infty$ & $\infty$ &$\infty$& finite & strong & strong\\
I$_l$  & Little-Rip (LR) & $\infty$ &$\infty$ & $\infty$ & $\infty$ & $\infty$ & finite & strong & strong\\
I$_p$  & Pseudo-Rip (PR) & $\infty$ &$\infty$ & finite & finite & finite & finite & weak & weak\\
II  & Sudden Future (SFS) & $t_s$ & $a_s$ & $\varrho_s$ & $\infty$ & $\infty$ & finite & weak & weak\\
II$_g$  &Gen. Sudden Future (GSFS) & $t_s$ &$a_s$ & $\varrho_s$ & $p_s$ &$\infty$& finite & weak & weak\\
III  & Finite  Scale  Factor (FSF) & $t_s$ &$a_s$ & $\infty$ & $\infty$ &$\infty$& finite& weak & strong\\
IV & Big-Separation  (BS) & $t_s$ &$a_s$ & $0$ & $0$ &$\infty$& $\infty$ & weak & weak\\
V & w-singularity (w) & $t_s$ &$a_s$ & $0$ & $0$ &$0$& $\infty$& weak & weak
\\
\hline
\end{tabular}
\end{center}
\end{table*}

Bearing in mind spacetime averaging (\ref{aver}) and the energy conditions (\ref{strong0})-(\ref{dom}), one can conclude that for BB singularities all the energy conditions are fulfilled, but averages vanish. For BR singularities no energy condition is fulfilled, but spacetime averages blow up. On the other hand, for an SFS, only the dominant energy condition is violated and averages are finite. It is interesting that on the ground of averaging, a BR singularity is a stronger singularity than a BB, while an SFS is weaker, and FSF is not necessarily so. Then, the averaging (\ref{aver}) seems to be a new kind of a measure for the strength of singularities \cite{PLB2011}.

\section{Varying constants removing or changing singularities}
\setcounter{equation}{0}
\label{varying}

The table \ref{classif} represents a variety of cosmological singularities -- each of them has different properties and they really appear in various physical theories such as superstring, brane, scalar field, alternative gravities and others. It seems reasonable that these singularities can be influenced by some processes which take place in the universe. For example, quantum effects add some extra terms which are higher-order in curvature \cite{houndjo}, and they may {\it change the strength} of singularities (e.g.  an SFS can be changed into either an FSF, a BR or a BB). So, one may ask the question if one is able to {\it remove or change the nature of singularities} using some other physical effects. One of the options is to make use of the time-evolution of the physical constants represented by appropriate scalar fields coupled to the gravitational field in the universe.

Having a look onto the generalized Einstein-Friedmann equations within the framework of the varying speed of light $c(t)$ (VSL) and varying gravitational constant $G(t)$ (VG) theories \cite{VSL}
\bea \label{rho} \varrho(t) &=& \frac{3}{8\pi G(t)}
\left(\frac{\dot{a}^2}{a^2} + \frac{kc^2(t)}{a^2}
\right)~,\\
\label{p} p(t) &=& - \frac{c^2(t)}{8\pi G(t)} \left(2 \frac{\ddot{a}}{a} + \frac{\dot{a}^2}{a^2} + \frac{kc^2(t)}{a^2} \right)~,
\eea
one is able to easily see that both the mass density $\rho$ and the pressure $p$ must be affected when $c$ and $G$ are varying. For example, if $\dot{a} \to \infty$, but $G(t)$ tends to infinity faster than $\dot{a}$, then a singularity in $p(t)$ can be removed.

For a flat ($k=0$) Friedmann model it is possible to write down an explicit relation between the pressure $p$ and the mass/energy density$\rho/\varepsilon$, though with a time-dependent barotropic index, in the form
\be
p(t) = w(t) \varepsilon(t) = w(t) \varrho(t) c^2(t)~~,
\label{pete}
\ee
where
\be
w(t) = \frac{1}{3} \left[2q(t) - 1 \right]~~,
\label{wute}
\ee
and $q(t) = - \ddot{a}a/\dot{a}^2$ is the (dimensionless) deceleration parameter.

It is worth mentioning that in view of the variation of the velocity of light $c(t)$ there is a crucial difference between the mass density $\varrho$ and the energy density $\varepsilon = \varrho c^2$, since variation of $c(t)$ effects the Einstein mass energy formula $E=mc^2$ transformed here as the
the mass density versus pressure formula $p = \varrho c^2$ after dividing both sides by the volume. Then, if one takes into account a barotropic equation of state, then
it is better to define the barotropic index which is dimensionless, as we did in (\ref{wute}). Since the pressure has the same units as the energy density, and the energy density results in multiplying the mass density by $c^2(t)$, then it is more reasonable to talk about singularities in the mass density and the pressure rather than in the energy density and pressure since they are, in fact, equivalent. The only factor which relates them is the barotropic index which we have assumed to be dimensionless. In other words, the power to remove a singularity by varying speed of light $c=c(t)$ refers only to the pressure, and not to the mass density. This can be seen from Eqs. (\ref{rho})-(\ref{p}). Taking into account any explicit form of matter such as for example the radiation $p(t) = (1/3) \varrho c^2(t)$ one easily sees that regularization with $c^2(t)$ can be done for the pressure only. The mass density singularity cannot be removed this way.

Effectively, impact of the variability of physical constants can be studied by the application of the scale factor, which admits a Big-Bang, a Big-Rip, a Sudden Future, a Finite Scale Factor, and a $w$-singularity
\be \label{newscalef} a(t) = a_s \left( \frac{t}{t_s} \right)^m \exp{\left( 1 - \frac{t}{t_s} \right)^n} ~, \ee
with the constants $t_s, a_s, m, n$ selected accordingly \cite{regular}. The first and the second derivatives of the scale factor (\ref{newscalef}) are
\bea
\label{dotnew}
\dot{a}(t) &=& a(t) \left[ \frac{m}{t} - \frac{n}{t_s} \left( 1 - \frac{t}{t_s} \right)^{n-1} \right] \propto \rho,~\\
\ddot{a}(t) &=& \dot{a}(t) \left[ \frac{m}{t} - \frac{n}{t_s} \left( 1 - \frac{t}{t_s} \right)^{n-1} \right] \nonumber \\
\label{ddotnew}
&+& a(t) \left[ -\frac{m}{t^2} + \frac{n(n-1)}{t^2_s} \left( 1 - \frac{t}{t_s} \right)^{n-2} \right] \propto p.~
\eea
From (\ref{dotnew})-(\ref{ddotnew}), one can see that for $1 < n < 2$ $\dot{a}(0) = \infty$ and $\dot{a}(t_s) = ma_s/t_s =$const., while $a(t_s) = a_s$, $\ddot{a}(0) = \infty$ and $\ddot{a}(t_s) = - \infty$ ($p \to \infty$), and we have an SFS.

\subsection{Removing a Big-Bang singularity -- VG}

Looking at the Eqs. (\ref{rho}) and (\ref{p}) we can see that the time-variation of the gravitational constant in the form
\be
\label{GBB}
G(t) \propto \frac{1}{t^2}~~,
\ee
(which is a faster decrease that in the early Dirac's ansatz $G(t) \propto 1/t$ \cite{dirac,alpha}), removes a Big-Bang singularity
in Friedmann cosmology (i.e. removes both $p$ and $\varrho$ singularities). In fact, in the Dirac's case $G(t) \propto 1/t$, only the $\varrho$ singularity can be removed. Besides, a time-dependence of $G=1/t^2$ is less influenced by the geophysical constraints on the temperature of the Earth \cite{Teller}.

Another suggestion is that the scale factor (\ref{newscalef}) would not approach zero at $t \to 0$ if it was rescaled be a ``regularizing'' factor $a_{rg} = (1+ 1/t^m)$ ($m \geq 0$), i.e.,
\be
a_{sm} = \left( 1 + \frac{1}{t^m} \right) \left(\frac{t}{t_s} \right)^m = \left(\frac{t}{t_s} \right)^m + \frac{1}{t^m_s}~~.
\ee

\subsection{Removing SFS or FSF - VSL}

It is also clear that any change in the mass density when the speed of light $c$ varies cannot be made without admitting a curvature term in the Einstein equations (\ref{rho})-(\ref{p}). This, especially refers to a Big-Bang singularity - it cannot be removed at all, unless the spatial curvature is non-zero. However, it is possible to regularize an SFS singularity by varying speed of light $c(t)$ if the time-dependence of the speed of light is given by
\be
\label{regc}
c(t) = c_0 \left( 1 - \frac{t}{t_s} \right)^{\frac{p}{2}}
\ee
($c_0=$ const., $p=$ const.), provided that
\be
p > 2 - n~~\hspace{0.5cm} (1<n<2)~~.
\ee
This, however, has an interesting physical consequence. Namely, the speed of light gradually diminishes reaching zero at the singularity. In other words, the {\it light slows and eventually stops} at an SFS singularity. This is surprising, although such an effect was already predicted within the framework of loop quantum cosmology (LQC).  In its anti-newtonian limit $c = \sqrt{1-2 \varrho/\varrho_c} \to 0$ for $\varrho \to \varrho_c$, with $\varrho_c$ being the critical density \cite{mielczarek}. In its  low-energy limit $\varrho \ll \varrho_0$ gives the standard value $c \to 1$. The effect also appears naturally in Magueijo model \cite{Magueijo01}, in which black holes are not reachable since the {\it light stops at the horizon} (despite they possess Schwarzschild singularity). An observer cannot reach the horizon surface even in his finite proper time. Interestingly, both limits $c \to 0$ and $c \to \infty$ are possible in Magueijo model.

One of the standard assumptions on the variation of the speed of light is that it follows the evolution of the scale factor \cite{VSL}
\be
\label{cc0}
c(t) = c_0 a^s(t)~~,
\ee
with $c_0$ and $s$ constant. The field equations (\ref{rho}) and (\ref{p}) become
\bea
\label{rhof}
\varrho(t) &=& \frac{3}{8\pi G(t)} \left(\frac{\dot{a}^2}{a^2} + kc_0^2 a^{2(s-1)} \right)~,\\
\label{pf} p(t) &=& - \frac{c_0^2 a^{2s}}{8\pi G(t)} \left(2 \frac{\ddot{a}}{a} + \frac{\dot{a}^2}{a^2} + kc_0^2a^{2(s-1)} \right)~.
\eea
With the time-dependence of $c(t)$ as in (\ref{cc0}), and with $a(t) = t^m$, it is possible to remove a pressure singularity provided $s > 1/m$ for $k=0$, $m>0$, and $s > 1/2$ or for $m<0$, $s < 1/2$ for $k\neq 0$, but not to remove the mass density singularity.

\subsection{Removing SFS or FSF - VG}

As it was already mentioned since for $k=0$ the mass density $\varrho(t)$ does not depend on $c(t)$ (see (\ref{rho})), then it is impossible to change an SFS singularity to become an FSF singularity. It is possible only with the gravitational constant $G$ evolving in time. Let us then assume that
\be
\label{regG}
G(t) = G_0 \left( 1 - \frac{t}{t_s} \right)^{-r}~~,
\ee
($r=$ const., $G_0=$ const.) which changes (\ref{rho}) and (\ref{p}) to the form
\bea
\label{rhoG}
\varrho(t) &=& \frac{3}{8\pi G_0} \left[\frac{m^2}{t^2} \left( 1 - \frac{t}{t_s} \right)^r - \frac{2mn}{tt_s} \left( 1 - \frac{t}{t_s} \right)^{r+n-1} \right. \nonumber \\
&+& \left. \frac{n^2}{t^2_s} \left( 1 - \frac{t}{t_s} \right)^{r+2n-2} \right]~,\\
\label{pG}
p(t) &=& - \frac{c^2}{8\pi G_0} \left[ \frac{m(3m-2)}{t^2}\left( 1 - \frac{t}{t_s} \right)^r \right. \nonumber \\
&-& \left. 6 \frac{mn}{tt_s} \left( 1 - \frac{t}{t_s} \right)^{r+n-1} + 3 \frac{n^2}{t^2_s} \left( 1 - \frac{t}{t_s} \right)^{r + 2n -2} \right. \nonumber \\
&+& \left. 2 \frac{n(n-1)}{t^2_s} \left( 1 - \frac{t}{t_s} \right)^{r+n-2} \right]~.
\eea
From (\ref{rhoG}) and (\ref{pG}) follows that an SFS singularity $(1<n<2)$ can be removed by varying gravitational constant when
\be
\label{con1}
r> 2-n~~,
\ee
and an FSF singularity $(0<1<n)$ can be removed when
\be
\label{con2}
r> 1-n~~.
\ee
On the other hand, assuming that we have an SFS singularity and that
\be
-1<r<0~,
\ee
we get that varying $G$ may change an SFS singularity onto a stronger FSF singularity for
\be
0 < r+n < 1~.
\ee
A physical consequence of the functional dependence of the gravitational constant in (\ref{regG}) is that {\it the strength of gravity becomes infinite} at a singularity. It is pretty obvious because we need to {\it overcome infinite (anti-)tidal forces} at the singularity. However, we now face another singularity - {\it a singularity of strong coupling} \cite{strongG} for a physical field such as for example the Brans-Dicke field $\Phi \propto 1/G$ \cite{bd}. On the other hand, physical field singularities were already dealt with in superstring and brane cosmology, where both the curvature singularity and a strong coupling singularity appeared (choice of coupling, quantum corrections \cite{PBB}).

\subsection{A hybrid case - VG}

A hybrid case which influences both types of singularities (Big-Bang and non-Big-Bang) would be
\be
G(t) = \frac{G_0}{t^2} \left( 1 - \frac{t}{t_s} \right)^{-r}~.
\ee
This ansatz would remove a big-bang singularity at $t=0$ with no additional conditions, while SFS or FSF singularities would be
regularized under the conditions (\ref{con1}) and (\ref{con2}), appropriately.

\section{Singularities and the limits of cosmology}
\setcounter{equation}{0}
\label{conclusion}

It is commonly thought that the {\it singularities are signs of the new physics} to be looked for and so are the milestones to limits of cosmology. However, nowadays one is able to differentiate quite a number of cosmological singularities which do not necessarily limit the physics, i.e. many of them are geodesically complete and can be ``gone through'' despite that they still lead to a blow-up of physical quantities (the scale factor, the mass density, the pressure, the physical fields etc.) as well as the mathematical entities (e.g. the curvature tensor). Some of the new singularities (which we call non-Big-Bang type here) {\it may serve as dark energy} in the sense that their appearance is related to physical theories such as string, brane, scalar fields, higher-order gravity $f(R)$, loop quantum cosmology etc. They can be fitted to observational data combined of supernovae, cosmic microwave background, baryon acoustic oscillations, and mimic standard $\Lambda$CDM model in redshift drift effect for specific choice of the parameters. Besides, they can be influenced by variability of physical constants treated as new physical (scalar) fields. In this context one shows that it is possible to remove or change the type of these singularities with the full physical consequences. The problem is that after a removal or a change one may face a new "singularity" of a physical field there which is responsible for the variability of constants. However, this is what happens in physical theories (e.g. superstring), too.

Some singularities (a Big-Bang, a Big-Rip) {\it are the limits of cosmology} (cannot be ``gone through'') {\it but some are not} despite admitting some "pathologies" (infinities) which in fact are not dangerous for the evolution of the universe.

The variation of the physical constants which influences the nature of singularities may be useful in the discussions of the {\it multiverse concept} giving the link through a kind of ``fake'' singularities to various parts of the universe with different physics which mark the limits of cosmology.

\section{Acknowledgements}

This work was supported by the National Science Center Grant DEC-2012/06/A/ST2/00395.

\end{document}